# Learning and Complexity in Genetic Auto-Adaptive Systems

Chris Adami

W.K. Kellogg Radiation Laboratory 106-38
California Institute of Technology
Pasadena, CA 91125

We describe and investigate the learning capabilities displayed by a population of self-replicating segments of computer code subject to random mutation: the tierra environment. We find that learning is achieved through phase transitions that adapt the population to whichever environment it encounters, with a learning rate characterized by the environmental variables. Our results suggest that most effective learning is achieved close to the edge of chaos.

# 1 Introduction

Our concept of learning, in artificial as well as natural systems, despite a plethora of instances, applications, and model systems, has remained intuitive. Indeed, there is as yet no general theory of learning (except for very specific systems [1]) and this omission is apt to become more and more crucial as experiments in learning become more and more varied and diverse. One of the more elusive tasks associated with formulating a theory of learning is the isolation of *universal characteristics* of the learning process. In fact, the very existence of a universal learning process has yet to be established.

In this paper, we would like to shed some light on the learning process in a very specialized artificial system that nevertheless promises to exhibit universal features. Also, the system offers the possibility to study learning from a biological, i.e. evolutionary, point of view. Evolution of DNA is perhaps the most dazzling instance of learning through adaptation that we know of. Yet, it seems to be of little use for machine learning applications for a very obvious reason: Learning through evolution is inherently slow. We hope nevertheless that by studying this immensely successful adaptive process, new insights can be gained which can be carried over to artificial learning systems.

In the next section, we would like to point out the qualitative differences between evolutionary learning (as displayed by natural genetic systems) and a variety of popular adaptive schemes that are in use today from an abstract point of view. The classification of learning processes introduced there is important for those readers interested in the conceptual foundations of learning, but may be skipped by those only interested in the results. Section 3 introduces the tierra system that serves as a paradigm throughout this paper, while the fourth section rigorously defines observables in tierra and introduces the equations that describe population kinetics. Section 5 then describes universal characteristics of the tierra system emerging from extensive simulations. We describe a typical tierra "experiment" in some detail and present results of an investigation of the learning rate as a function of the external mutation rate, i.e., the force that drives evolution. We offer conclusions in the last section.

# 2 Learning in Adaptive Systems

In learning, we are interested in the macroscopic behaviour of a system in response to external stimuli. When specifying the macroscopic state of the system, we are faced with two possibilities: We may either specify the *space* of macroscopic states by enumeration (i.e., providing each state fully formed), or else provide a set of *microscopic* states together with a set of rules to construct the macroscopic ones. Either of these approaches has its advantages. The macroscopic implementation is well suited for complex tasks to be learned as each preprogrammed state can in principle be of arbitrary complexity. On the other hand, as will become clear later, flexibility is lost and the set of possible states is necessarily finite. The microscopic approach does not suffer from the latter problem because the microscopic rules can be combined in an infinite number of ways to produce a practically infinite set of macroscopic states. Providing a "microscopic alphabet", however, in which every macroscopic rule can be formulated, seems daunting most notably due to the hierarchy problem and the brittleness problem.

The hierarchy problem is most easily understood by considering its analogue in natural language: the parsing problem. In natural language, the meaning of a sentence can *not* be a universal function of the words, simply because words have no *intrinsic* meaning at all. Rather, the meaning of a word is given by all the possible ways it can be used in a meaningful sentence. Thus, there is no meaning on the microscopic level, whereas we need a meaning on the macroscopic level. The mapping between the levels cannot be performed by a universal function because words are universal (the same set of words are used to construct all sentences) while the sentences are not (the meaning of sentences is context-specific). In learning systems, fitness replaces meaning, microscopic states (the alphabet) replace words, and macroscopic states (the rules) replace sentences. The alphabet must be devoid of intrinsic fitness in order to guarantee universality, i.e., the fitness of a certain arrangement of the microscopic states should not be a universal function of the fitness of each member of the alphabet, while we would like to see fitness emerge (on the macroscopic level) that is inherent to the context and thus *nonuniversal*, and only reflects the properties of the environment, i.e. the learning task at hand.

The brittleness problem is well-known: an arbitrary arrangement of microscopic rules leads to nonsensical macroscopic rules in almost all cases, and the space of macroscopic states turns out to be mostly empty. This problem most notably arises with computer-code for von Neumann machines: the ratio of possible programs to workable ones is almost zero, and any arbitrary mutation of a working program will most likely break it.

As a consequence of these problems, most approaches to the learning problem are based on the macroscopic implementation. Here, the major players in the field are Artificial Neural Networks [2], Genetic Algorithms [3, 4] (including Expert Systems), and certainly Kauffman's NK-model [5]. All these are instances of "adaptive" systems, which learn by adapting to the fitness landscape dictated by the task to be learned. They share the ubiquitous feature that is the feedback mechanism: a process which modifies parameters that determine the response of the system to a certain input, according to the fitness, or success-rate, of the previous set of parameters. In conventional adaptive systems, the mechanism to determine the fitness of a parameter set is extraneous to the system itself. This is of course a direct consequence of the inability to provide a microscopic problem-independent alphabet, as the parameter-string (or set of weights and thresholds) has *no significance* except when interpreted within the context of the fitness-function or error-function. Thus, the system can never learn anything outside the boundaries specified by this function: flexibility is lost. As it turns out, Nature seems to have found a solution to this problem, and we attempt to emulate this approach.

In almost all cases of learning in natural systems, the fitness of a certain configuration (or "hypothesis" [1]) is determined *within* the system. Thus, we strive for the fitness of a string (in the broad sense of NK-models) to emerge as a collective effect from the interaction of the environment (a "hard-coded" set of parameters) and the population. In a way, we would like the strings to *compute their own fitness*. We shall call systems that can perform this feat "auto-adaptive", to emphasize the fact that we *do not provide a fitness- or*

*error-function.*

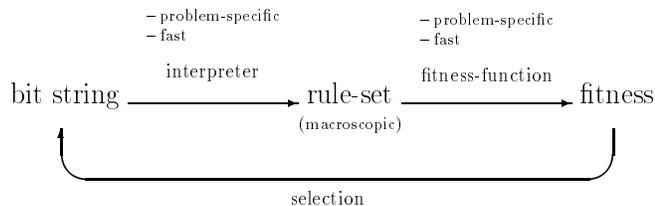

Fig. 1a

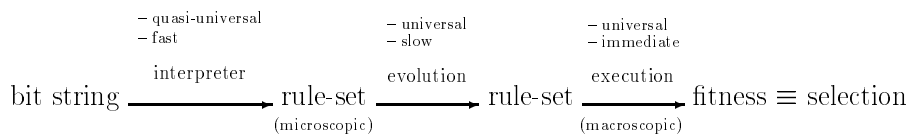

Fig. 1b

Fig. 1. (a) Feedback loop for adaptive genetic systems for selection of strings. (b) Selection process for auto-adaptive genetic systems.

Fig. 1 is an attempt at schematizing adaptive and auto-adaptive systems. We assume that the information content of any learning system may be coded in bit strings. In adaptive systems (Fig. 1a), the bit strings are translated into macroscopic sets of rules[1]. This interpreter is necessarily problem-specific, and the construction of the rules (the action of the interpreter) is *fast* (on the time scale associated with the learning process). The fitness of this macroscopic rule-set is then computed via the external fitness-function, which is also problem-specific, and fast. The result of the fitness-evaluation is used to select bit strings in the next generation. The bit strings of auto-adaptive systems (Fig. 1b) are first translated to a microscopic rule-set. This interpreter is quasi-universal: the same microscopic rule-set can in general be used for any application, although it may in most cases turn out to be advantageous to adapt the interpreter to a specific *class* of problems. The action of this interpreter is fast. The assembly of microscopic rules to macroscopic ones proceeds via mutation, evolution, and "natural" selection[2]. This process is universal, but slow on the time scale of generations. The fitness evaluation then does not require any more manipulation. Instead, the fitness emerges through the (social or non-social) interaction of the macroscopic rule-sets in the population. Thus, fitness is the direct result of the actions and interactions of the members of the population, and is automatically the vehicle for selection of bit strings that survive in the next generation. For this to work, inevitably the bit strings have to self-replicate.

---
[1] We use the terms 'rule-sets' and 'states' synonymously, as each state of a system can in fact be viewed as a set of rules to handle input and output.

The only (artificial) system, that (to our knowledge) is truly auto-adaptive was designed to mimic nature in a number of important aspects. The tierra environment [6], a software package created recently by Tom Ray, an evolutionary biologist, is one where a population of self-replicating segments of computer code (alternatively called 'programs', 'cells', or 'creatures') thrives in an environment that is managed by the tierra program itself. The latter provides not only resources to the cells (CPU-time and memory-space), but also oversees births, mutations, and deaths, along with providing the "shells" in which the creatures live: a virtual computer for each living cell in the population. Before we go on to describe the key aspects of the tierra system, we would like to clarify the recurrent use of metaphors culled from biology. In fact, tierra was designed around these metaphors, in the sense that certain devices of the computing environment were designed to *play the same role* as certain devices, in the broadest sense, occurring in nature. Thus, CPU-time is *analogous* to energy, memory-allocation is *analogous* to birth, machine-language instructions (the microscopic rule-set) are *analogous* to the codons of DNA[3].

---

[2]We put "natural" in quotes since the selection process is of course dictated by the user-specified environment. However, we would still like to use the term "natural" to distinguish it from "artificial selection" based on the output of a fitness-function. More accurate terms would be "internal" as opposed to "external" selection.

[3]DNA is coded in base 4 deoxyribonucleotides, such that any sequence of 3 represents a codon that is translated into an amino acid (the microscopic rule-set of nature). Thus, $4^3$ codons are translated into 20 amino acids, while $2^5$ combinations of 1's and 0's are translated into 32 instructions in tierra , some of which turn out to be rarely used and could

It turns out *a posteriori* that such a system of analogies and metaphors can, to an extent dictated by hardware limitations, emulate the evolution of simple proto-cellular systems to an astonishing degree [6].

The replication and mating operations, extraneous to the population of strings in Genetic Algorithms (GA's) for example, is inherent to the tierra community of cells and as such the control of these activities is shared between the environment and the make-up of the population. Giving up control over key parameters has profound consequences for the macroscopic behaviour of the population. Loss of microscopic predictability increases the complexity of the system to such a degree that studies of the tierra system are in effect experiments with tierra . Concurrently, complexity ensures that the collective behaviour of the population is genuine and reproducible, and, in its general characteristics, universal.

# 3 The tierra System

The notion to evolve computer programs by means of random mutation appears doomed owing to the fact that the ratio of working programs to possible ones is very close to zero for most existing languages. In other words, any random mutation of a program is likely to break it. This has been known for some time as the problem of "brittleness". On the other hand, mutation does quite well in living systems, and according to Darwinian theory, is responsible for the emergence of complexity in natural living systems. Ray dissolved this dichotomy by designing an assembly language based on a number of instructions of the same order of magnitude as there are amino acids in the genetic code. Specifically, he chose to code these instructions into five bits, such that the random mutation of any bit would be "contained", and lead to a different instruction of this family. This is the key idea to surmount brittleness, and possibly the key to auto-adaptive systems in general. Another characteristic of the tierran instruction set garnered from nature is the use of

---

just as well be eliminated for redundancy. In DNA, those amino acids that are used most frequently have the most representations in terms of codons. Such an approach can easily be implemented in tierra also.

templates (patterns of instructions) for addressing purposes rather than absolute addresses. From a computing point of view, the 32 instructions used by the tierran creatures are similar to machine language instructions; an extremely reduced instruction set running on the virtual computers provided by the tierra program. The virtual CPU is kept very simple using four registers, a stack, input/output buffers, and an instruction pointer. Table 1 shows the mapping from the tierran codons to instructions.

| 00000 nop0   | 01000 pushdx | 10000 dec   | 11000 jmp    |
| 00001 nop1   | 01001 popax  | 10001 add   | 11001 jmpb   |
| 00010 movdi  | 01010 popbx  | 10010 sub   | 11010 call   |
| 00011 movid  | 01011 popcx  | 10011 zero  | 11011 adr    |
| 00100 movii  | 01100 popdx  | 10100 shl   | 11100 adrb   |
| 00101 pushax | 01101 put    | 10101 not0  | 11101 adrf   |
| 00110 pushbx | 01110 get    | 10110 ifz   | 11110 mal    |
| 00111 pushcx | 01111 inc    | 10111 iffl  | 11111 divide |

**Tab. 1** Mapping of 5-bit codons to instructions in the tierran instruction set used in the present simulations. A description of the commands can be found in the tierra manual [8].

The intended analogy is for the strands of computer code to represent strands of DNA, while the tierra program fulfills the role that chemistry plays in nature. Specifically, it doles out CPU time-slices to the cells in the group (simulating parallel coexistence) and supervises the "aging" of the cells by arranging them in a "reaper queue", killing the oldest cells in the strip of memory reserved for the cells (the "soup") if there is not enough room to accommodate the new-born ones. Details of the operation of the queues and the observing software which is part of the tierra program can be found in [6] and in the documentation of the tierra software [8].

Evolution of the population is guaranteed by a rate of bit mutation that affects every cell in the soup to the same degree (this is the analog of cosmic rays). Mutations in the cells due to this phenomenon and to random copy-errors seems to be the key mechanism that drives the emergence of complexity, learning, and diversity. The "splicing" mechanism of mating which is the corner stone of the evolution of GA's arises in tierra as a secondary effect of mutation and flaws, by copying an incomplete creature (incomplete due to a mistake in calculating the cells' length as a result of mutation and flaws) into the space previously held by a now defunct one, thus splicing these codes together. It turns out that this mechanism plays an important role in learning and the evolution of complexity on short time scales. Also, it is an example of an emergent characteristic, as it was not even remotely anticipated by the designer [9].

A typical tierra experiment starts by inoculating empty memory by a self-replicating creature that is hand-written by the operator using any suitable instruction set. Throughout, we inoculate the soup with our equivalent of a program written and termed "the ancestor" by Ray[4]. The ancestor is a code consisting of 82 instructions that represent Ray's first attempt at writing a self-replicating program for this particular instruction set. As such, it turns out to be very inefficient and is easily improved by mutation. We use it as a progenitor for precisely this reason, as its inefficiency is equivalent to the presence of redundancy in the code. Redundancy has emerged as a necessary requirement for successful evolution. Also, this progenitor possesses only the ability to replicate, and thus is not biased towards learning other tasks. After inoculation, the reserved space for the cells quickly fills up with offsprings of the ancestor, largely identical to it, with exceptions due to mutations. Once the space is filled up, the tierra program reaps the oldest cells to provide room for the next generation. As mentioned, age is controlled by arranging the cells in a linear queue. New-born cells are entered at the bottom while the top creature is removed. From the moment of inoculation, the fate of the population is out of the hands of the operator, being entirely determined by the parameters of the tierra program and the physical environment (the "landscape") encountered by the cells (see below). Despite the evidently deterministic relationship between parameters and macroscopic behaviour, the system is complex enough

---

[4]Our ancestor is not exactly identical to Ray's due to some slight changes in the instruction set that we deemed advantageous. The instruction set used in the simulations here is displayed in Table 1.

to thwart any attempt at unraveling that connection.

## 4 Fitness and Learning in tierra

As mentioned in the previous section, the fitness of a member of the tierran population is *not* determined by a fitness-computation, but rather is a function of the cells genotype[5] *and* of the rest of the population. A universal measure of fitness in tierra, as well as possibly all auto-adaptive systems, artificial and in a restricted sense natural, is the number of off-spring ("daughters") of the organism $i$, $d_i$, in a suitably chosen time span. In tierra, we take this span to be the lifetime of the organism, $\tau_i$, measured in number of instructions executed. Very obviously, in the absence of a mechanism that allows organisms to kill each other, the genotype with the highest number of off-spring per lifetime will dominate the population. Naturally, this dominance can only be ephemeral as the successful creatures' off-spring will soon compete with it.

The number of daughters (during its lifetime) of organism $i$ can be written as

$$d_i = \frac{\tau_i}{(t_g)_i} \tag{1}$$

where $(t_g)_i$ is the time it takes organism $i$ to gestate a single off-spring, and $\tau_i$ is the lifetime of this organism as defined earlier. In the emulation of parallel coexistence, the main program allocates slices of CPU time to each cell in a serial manner. Let $(t_a)_i$ be the time allocated to organism $i$ (measured in number of instructions that this cell will be able to execute) in each sweep through the population. Then

$$\tau_i = \sum_{j=1}^{N_i} (t_a)_{i,j} \tag{2}$$

where $N_i$ is the number of sweeps that creature $i$ obtains. Let us for simplicity also assume that the time allocated each sweep is roughly equal (or equivalently

---

[5]The genotype of a cell is given by its specific arrangement of instructions. For programs of the same length, different genotypes are arbitrarily labelled by a three-letter code, in order of their appearance in the soup. Thus, the size-82 progenitor is labelled 82aaa, its first off-spring of the same size with a different genotype is 82aab, and so forth.

define $(t_a)_i$ to be the average allocated time per sweep). Then

$$\tau_i = N_i(t_a)_i \, . \tag{3}$$

It then follows that (we will in the following drop the subscript $i$ denoting the value of the respective quantity for organism $i$, while denoting quantities averaged over the entire soup by angled brackets)

$$d = N\frac{t_a}{t_g} \equiv N\alpha \tag{4}$$

where we defined the *fitness fraction* $\alpha$.

Indeed, this fraction is a function of the genotype of the organism only, and thus represents a good measure of *absolute fitness*. The total number of off-spring $d$ can only be a measure of relative fitness as its value depends on the number of off-spring of other members of the population through its dependence on $N$. A good estimate for $N$ is obtained by considering the movements in the Reaper Queue (RQ) due to new births only. As mentioned briefly earlier, every new-born cell is entered at the bottom of the queue, and reaches the top after $n$ more births, where $n$ is the total number of cells in the soup. The oldest cell in the soup is the one at the top of the queue, and suffers the action of the reaper. Since $\langle\alpha\rangle n$ is the average number of cells born each sweep, a constant population implies $N\langle\alpha\rangle n = n$ and thus

$$N = \frac{1}{\langle\alpha\rangle} \, . \tag{5}$$

It then follows that

$$d = \frac{\alpha}{\langle\alpha\rangle} \, . \tag{6}$$

In tierra however, there is also movement in the RQ which is not due to births and deaths alone. If a cell attempts an illicit operation, be it writing on write-protected memory space (for instance space owned by another creature), or attempting to allocate too much or too little memory[6], an error-flag is set, and the instruction is not executed. Anytime a cell obtains a time-slice (thus every sweep), its total number of error-flags $n_e$ is compared to the number of error-flags generated by the cell just above it in the RQ, and switches places with it if that cells error-count is larger. Thus, cells that commit more error-flags age faster. On the same token, a cell may be moved *down* the RQ if it accomplishes a task that the user feels worth rewarding. In the present implementation of tierra, a cell moves down one position in the RQ after a successful memory allocation instruction (`mal`), and after a successful `divide` instruction. The number $k$ of downward moves per lifetime ($k = 2d$ in the task-neutral case) is at the discretion of the user and represents a means of rewarding or punishing cells according to whatever task is to be accomplished. Including these movements inside the RQ, we find the more general expression for the number of off-spring

$$d = \frac{\alpha + \alpha/n\left(\langle n_e \rangle - n_e\right)}{\langle \alpha \rangle + k/n \left(\frac{\langle \alpha \rangle n_e - \alpha \langle n_e \rangle}{n_e + \langle n_e \rangle}\right)} \ . \tag{7}$$

---

[6]The do or don'ts are set by parameters of the tierra software. See the documentation for the details.

Note that the corrections to (6) are of the order $1/n$, and thus become more and more unimportant in simulations with large $n$. This is due to the fact that the reaper kills the oldest cells in the entire soup, while a more sophisticated model would consider removing the oldest cell in a specific neighbourhood of $\tilde{n}$ cells [7].

Another method of rewarding some actions and discouraging others is the distribution of *bonuses* in the form of extra time-slices. For an organism of length $\ell$, tierra doles out slices of

$$t_a = (c + f)\ell^p + t_b \tag{8}$$

instructions per cell per sweep. Here, $t_b$ is the average bonus received per sweep, $p$ is a power that can be used to favour larger or smaller creatures (we set $p = 1$ for size neutrality throughout) and $f$ is the "lean-ness" fraction of the cell, obtained by dividing the number of executable instructions of the cell by its length. This factor is introduced to discourage the development of unexecutable code (as occurs if a section of the code is jumped over by the instruction pointer. This would be advantageous as it reduces the gestation time as we shall see below). We have supplemented this fraction by a genotype-independent constant $c$ ($c = 0.3$ throughout the simulations reported here). For comparison, the ancestor has $f = 0.54$, but evolution is able to increase this fraction to close to the theoretical maximum of $f = 1$.

Concurrently to increasing $t_a$, cells can decrease $t_g$ in order to increase $\alpha$. Let us divide a typical program into a "work" section of length $\ell_w$ and a copy loop of length $\ell_c$, such that $\ell = \ell_w + \ell_c$ (with typically $\ell_c \ll \ell_w$). The copy loop consists of those instructions that have to be executed to copy instructions from mother to daughter. Thus, to copy $\ell$ instructions, a total number of $\ell \ell_c / m$ instructions have to be executed, where $m$ is the number of instructions copied by executing the instructions in $\ell_c$. In the ancestor $m = 1$; however, the cells quickly discover that increasing $m$ reduces the gestation time. This technique of optimization is generally known as "unrolling the loop" and was observed to occur spontaneously in tierra by Ray [6]. To complete a gestation, the program also has to run through the remaining $\ell_w$ instructions, such that

$$t_g = \ell_w + \frac{\ell \ell_c}{m} = \ell \left( 1 + \ell_c \left[ \frac{1}{m} - \frac{1}{\ell} \right] \right) \tag{9}$$

and thus

$$\alpha = \frac{c + f + t_b/\ell}{1 + \ell_c(\frac{1}{m} - \frac{1}{\ell})} \ . \tag{10}$$

For $\ell_c \ll \ell$ and small $m$ ($m \lesssim 3$) we find $\alpha \sim m$, i.e. unrolling the loop is an extremely beneficial operation. For larger $m$ the lengthening of the copy loop cuts down this advantage. Likewise, skipping a large part of $\ell_w$ would turn out to increase $\alpha$ substantially. However, this is detrimental to learning as this is precisely the region where the cells are supposed to develop the code necessary to accomplish a task. For this reason, the lean-ness factor $f$ was introduced in (8) above.

The mechanism that drives fitness-improvement in tierra is of course mutation. The soup is subject to independent, Poisson-random mutation (i.e. bit-flip) events, such that the waiting times between mutations are distributed exponentially[7]. The mean time between mutations $\langle t_m \rangle$ is related to the mutation rate $R$ and the soup size $s$ via

$$\langle t_m \rangle = \frac{R^{-1}}{s} , \qquad (11)$$

while the probability that two mutation events are spaced by $t_m$ is

$$p(t_m) = Rs\, e^{-Rs\, t_m} = \frac{1}{\langle t_m \rangle} e^{-t_m/\langle t_m \rangle} . \qquad (12)$$

We are now in a position to obtain a relationship between the fitness of a genotype $i$, $\alpha_i$, and the mutation rate.

The number of cells of genotype $i$ in the soup at time $t+1$, $n_i(t+1)$, is related to $n_i(t)$ via

$$n_i(t+1) = \left(1 + \frac{\alpha_i - \langle \alpha \rangle}{t_s} - R\ell_i\right) n_i(t) . \qquad (13)$$

Eq. (13) simply reflects that new cells of genotype $i$ are born with a rate $\alpha_i/t_s$ ($t_s$ is the time it takes to "sweep" through the soup once, i.e. to execute $(t_a)_i$ instructions for each each cell in the soup, $t_s = n\langle t_a \rangle$) while the fitness $\alpha_i$ is just the number of off-spring per sweep) and they die with a rate $\langle \alpha \rangle/t_s$ due to births by other genotypes, and with a rate $R\ell_i$ due to mutations. We can neglect here the rate of births of this genotype due to mutations affecting the rest of the soup, as this is infinitesimal in most situations[8]. For simplicity, we also neglect in this equation the effect of mutations due to copy-errors, which enters in the first term of (13). For a copy-error rate $R_c$ (one out of $R_c^{-1}$ instructions are not copied correctly) the term $\alpha_i/t_s$ in (13) should be multiplied by $(1 - R_c \ell_i)$. In the present paper we set $R_c = 1 \times 10^{-3}$ such that it can safely be ignored at medium and high background-mutation rates.

Solving (13) we find for the evolution of the population

$$n_i(t) = n_i(t_0) e^{\gamma_i t} \qquad (14)$$

---

[7]This is an improvement over the univariate distribution in earlier versions of tierra .
[8]This term however is important in a consistent treatment of the statistical mechanics

where $n_i(t_0)$ is some starting population (e.g. $n_i(t_0) = 1$) and (suppressing the genotype-index)

$$\gamma = \frac{\alpha - \langle \alpha \rangle}{t_s} - R\ell \ . \qquad (15)$$

Likewise, this allows us to derive a relation for the maximum mutation rate that a population of fitness $\alpha$ can sustain. The highest strain is put on a population during a parasite invasion, where up to 90% of the species are eradicated and the average fitness of the soup is driven close to zero, $\langle \alpha \rangle \to 0$. Then the soup can only survive if the best genotype has $\gamma \geq 0$ [see (14)], or

$$\frac{\alpha}{t_s} \geq R\ell \ . \qquad (16)$$

In other words, there is a minimum fitness (i.e. minimum replication rate) required to survive under the hostile circumstances of a high mutation rate. This condition is similar to the error-threshold condition derived by Eigen et al. in the context of quasi-species in protein-space [10]. Averaging (16) over genotypes gives us a more intuitive understanding of the requirements for the survival of a population. Since $t_s = n \langle t_a \rangle$ we find

$$\frac{1}{\langle t_g \rangle} \geq R \langle \ell \rangle n = \frac{1}{\langle t_m^\star \rangle} \qquad (17)$$

where $\langle t_m^\star \rangle$ is the average time between mutations affecting cells (as not all sites in the soup are actual cell-sites), $t_m^\star = (s/n\langle \ell \rangle) t_m$.

The survival condition is thus a relationship between the two fundamental (small) time scales in the problem, the gestation time $\langle t_g \rangle$ and the average time between cell-mutations, $\langle t_m^\star \rangle$. Not surprisingly, we find that we must have

$$\langle t_g \rangle \leq \langle t_m^\star \rangle \ . \qquad (18)$$

While this equation was derived for the system considered under special conditions that may not hold in more realistic systems, we do expect such a relation to hold quite generally.

By the same token, Eq. (15) tells us how the mutation rate drives the fitness improvement. As equilibrium always drives any genotype towards $\gamma \approx 0$, we

find

$$\Delta\alpha \equiv \alpha - \langle\alpha\rangle = R\ell t_s ,\qquad(19)$$

i.e., the fitness gradient is proportional to the mutation rate. Of course, this equality is violated during the phase transitions that improve the fitness, i.e. during learning.

In order to gain some insight into how the mutation rate affects the *learning rate* (which after all is our prime focus in this paper), we need to perform actual experiments with tierra. We have seen that there is a maximum rate above which the soup cannot survive, while obviously there can be no learning at $R = 0$. We shall in fact see in the next section that, although a learning rate cannot unambiguously be linked to a mutation rate, they are in effect loosely correlated until close to the transition to chaos, which effectively dissolves the population: a state where self-replication stops.

## 5  Characteristics of Learning

In order to observe learning in tierra, we investigate a simple problem: learning to add two integer numbers. We choose this problem as a representative of a class of easy problems[9] that can be mastered by a tierran soup, anticipating that more complex problems can be learned by combining such microscopic tasks. Since the tierra system is a parallel one *in principle* (though not in practice), learning several tasks at once should not require the cumulative time of learning each of them.

As opposed to e.g. learning in Neural Networks, we do not "teach" the system using a certain set of data only to test it with a foreign one later on. Rather, we embed it in an environment that is biased towards a certain task, i.e., we *present it with the information that adding is advantageous*. Furthermore we provide numbers in the input buffer of each CPU that the tierran cells may choose to manipulate, but nothing more. While the cells eventually learn to add just these numbers, these may be exchanged with *any* other numbers at *any* given time. Thus, the cells truly learn the concept, not just an instance.

---
[9]An attempt at solving the XOR problem using tierra is described in Ref. [11].

Our main tool to bias evolution towards accomplishing the chosen task is the distribution of bonuses in the form of extra time [cf. Eq. (8)]. We reward three accomplishments which are formulated in as general a manner as possible so as not to bias towards any particular solution to the problem. The first step consists in rewarding cells that develop the correct input/output structure for the problem at hand. Clearly, adding requires a minimum of two inputs and one output. As a consequence, any cell that develops a minimum of two `get` and a minimum of one `put` command receives a certain bonus at the time of gestation of an off-spring (Table 2 lists the specific bonuses used in the simulation presented here). The next step is "clearing the channels", in other words we reward cells that manage to echo the values in the input buffers into the output buffers. Finally, any cell that writes a value into the output buffer that happens to be the sum of the two previously read values is rewarded with extra time at the time of executing the successful `put` command. Note that any such bonus increases the fitness of such a cell according to (10) resulting in more off-spring for that cell and a subsequent perpetuation of the discovery.

| feature | bonus |
|---:|---:|
| input/output | -50 |
| echo | 40 |
| add | 100 |

**Tab. 2** Distribution of bonus for evolved features. A negative bonus indicates that this number of instructions is *subtracted* from the default allocated time-slice if this feature is *not* evolved.

The rewards are of course available simultaneously and can in principal be discovered in any order. This reward-structure, "soft-coded" into the instruction set[10], constitutes the "fitness-landscape" with valleys, mountains, and ridges, that the soup has to adapt to in order to thrive.

---

[10] We distinguish between the "hard-coded" part of the instruction set which is the same ("universal") for any problem (and could just as well be etched into silicon) and the "soft-coded" part which is specific to the problem at hand, and thus represents part of the "physical" environment that the cells live in.

In the simulations presented here the environment is extremely simple, there being only three distinct explicit bonuses. However, they can be combined in different ways, and two of them can (in the present simulations) be repeated up to three times to gain additional bonus. Also, there is only a limited number of ways for a cell to reduce its gestation time (resulting in higher fitness). The introduction of the lean-ness factor $f$ on the other hand already provides for a means to improve fitness in a quasi-continuous way (up to $f = 1$). Furthermore, cells can exploit the structure of the population itself to gain fitness, a feat most impressively demonstrated by the parasites (sections of code that cannot reproduce on their own, but rather use the copy-loop of a host cell to produce off-spring). In all these instances of fitness-improvement, information is "found" by a cell (through mutation) and used to gain an advantage. This information is then reflected in the genome of the adapted cell.

Even though the environment for the adding problem is extremely simple, the space of possible fitness improvements appears to be extremely large. Since every genotype has a specific fitness, we can think of the space of possible fitnesses pertaining to the problem as meta-stable states in a continuum of fitness states while transitions between these states are driven by mutations. Since the number of meta-stable states is already very large for this simple example (and should effectively be infinite in any realistic system) the tierran system will exhibit features of a self-organized critical (SOC) system[11]. As a consequence, the time between transitions (or "avalanches" in the language of SOC's) is distributed according to a power law (as are the sizes of fitness-jumps) and thus an "average time between transitions" (which would allow a determination of the learning rate) cannot be defined.

As is well-known [12], a power-law distribution of waiting-times is due to an absence of scales in the problem. This is true to a certain extent in tierra as there is no time scale of the order of the time scale of learning (or evo-lution), nor is there a scale setting the size of the avalanches[12]. There are however microscopic time scales [those of Eq. (18)] and these lead to a violation of power-law behaviour. It is precisely the presence of these scales that leads to a correlation between $\langle t_m \rangle$ (the average time between mutations) and the learning rate. The latter is still notoriously difficult to define. One approach would be to determine the average time taken to learn the specified task at a fixed mutation rate, yet the measured times scatter heavily around the average due to the stochastic nature of the learning process. In principal there is no guarantee that in any specific simulation the goal will be attained (i.e., the learning time is in principle infinite) while in practice the goal is (at large enough mutation rates) almost always attained (see below). On the same token, it is impossible to predict the sequence of meta-stable states that the soup will traverse to reach the maximum fitness (pertaining to this problem). As a consequence, the end product (i.e., the most successful genotype) will very seldomly look the same even for two runs with exactly the same starting conditions (except for the random seed) and thus exactly the same "environment". This is strong evidence for contingency in the learning process for auto-adaptive genetic systems, and possibly for evolutionary processes in general.

---

[11] A full investigation of self-organized criticality in tierra is outside of the scope of this paper and will be reported elsewhere.

Fig. 2 shows the evolution of fitness (the learning curve) in a typical run at an intermediate mutation rate, specifically $R = 1.33 \times 10^{-8}$ with a soup-size $s = 131072$ which translates into an average time between mutations of $\langle t_m \rangle = 1/Rs \approx 572$ instructions. The upper curve depict[4 s the fitness $\alpha$ (defined in section 4) of the "best-of population" (the genotype with the most living copies) every million instructions, which translates roughly into every three generations. The lower curve shows the average fitness of the population $\langle \alpha \rangle$. As expected, the fitness-of-the-best increases via jumps indicating transitions between meta-stable states. These are most likely first-order phase transitions as is evident from the coexistent phases. (A detailed investigation of the statistical mechanics of this system will appear elsewhere).

---

[12]This is only approximately true in the simulations presented here, as the paucity of rewards (see Table 2) does set a scale at large fitness-jumps.

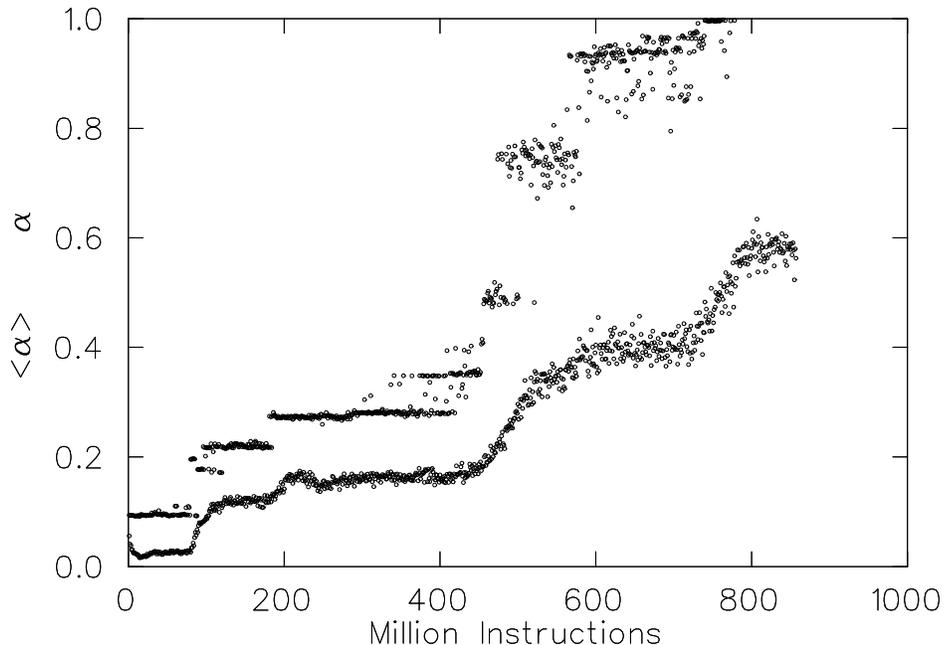

Fig. 2. Learning curve for a simulation with $R^{-1} = 75$ million instructions, in a soup of size 131072 instructions. The upper curve depicts the fitness $\alpha$ of the "best-of-population" while the lower one shows the average fitness $\langle \alpha \rangle$ of the population.

The first transition in Fig. 2 (at around $t = 80$ million executed instructions) is in fact due to the unrolling of the loop mentioned earlier, which literally halves the gestation time of the cell. Consequently, $\alpha$ jumps by roughly a factor 2. The transition at $t = 100M$ involves a minor rearrangement of code, while at $t = 185M$ the copy-loop is unrolled to $m = 3$. The input/output structure (first bonus in Tab. 2) is achieved around $t = 290M$ but appears as only a small increase in fitness. This is due to the bonus being distributed over several sweeps, which entails that the average gain per gestation-period is rather small. Echoing is learned at 340M (this time is defined as the time when a cell that discovered echoing dominates the population for the first time). The transition at $t = 409M$ simply makes echoing more efficient, and prepares the ground for the transition at $t = 453M$, when the best-of-population simultaneously triggers the bonus for adding *and* echoing. This is not a rare scenario, as cells often first develop the capacity to echo twice in a gestation period thus earning a bonus of 80, only to transform one of the echoing sections of code into an adding one. The later transitions simply accumulate echo's and add's (mainly by splicing together sections of code containing the pertinent sequence) so as to trigger the maximum bonus.

The complexity of the cell dominating the population at around $t = 800$M is intriguing. Not only has it evolved the capacity to successfully manipulate the numbers in the input buffers by adding them several times per gestation period, but it also optimized its reproduction loop to gestate off-spring three times faster than the ancestor[13]. While the cells will always attempt to do the latter, we could have rewarded an entirely different task, and consequently the final genotype would reflect that in its genome instead. In fact, after the cells learn to write the content of the input buffers into the output buffers, an inspection of the output buffers of all coexisting cells at that moment shows that all kinds of operations are performed on these numbers. The majority of the cells return the input-numbers untouched so as to trigger the 'echoing' reward, some however subtract them, add all three, subtract the number 4, and so forth. The reward structure simply weeds out those cells with mutations that allow them to add two numbers out of the zoo of creatures that perform a litany of tasks, entirely accidentally. In this sense, the actual nature of the task is irrelevant for the general characteristics of learning in the tierra system.

We have performed this type of experiment ten times for each of eight different mutation rates, at a constant soup size. The mutation rates were chosen to range from very low, where adding is achieved only very late (if at all), to very high rates where the population effectively "melts" (ceases to reproduce). This happens at around the point where $\langle t_g \rangle > \langle t_m^\star \rangle$ as derived earlier, i.e., when on average a cell is hit by a mutation before it can generate its first off-spring. Clearly then, a cell cannot on average propagate its genome, and the information contained in it. For each mutation rate, the learning time fluctuates strongly due to the statistical nature of the learning process and to the presence of meta-stable states in the system that can trap the population.

---

[13]With our current version of tierra, 800M instructions are reached on average after about 5 hours of CPU time on an HP 9000/750 workstation.

The time it takes for the population to escape such a trap then determines the learning time. In most such cases, we were unable to wait long enough to see this happen. Fig. 3 shows a learning curve for half the mutation rate in Fig. 2, where the population was stuck in a meta-stable level of fitness $\alpha \approx 0.2$, before breaking out of it at around $t = 1900M$ and learning to add almost instantly after.

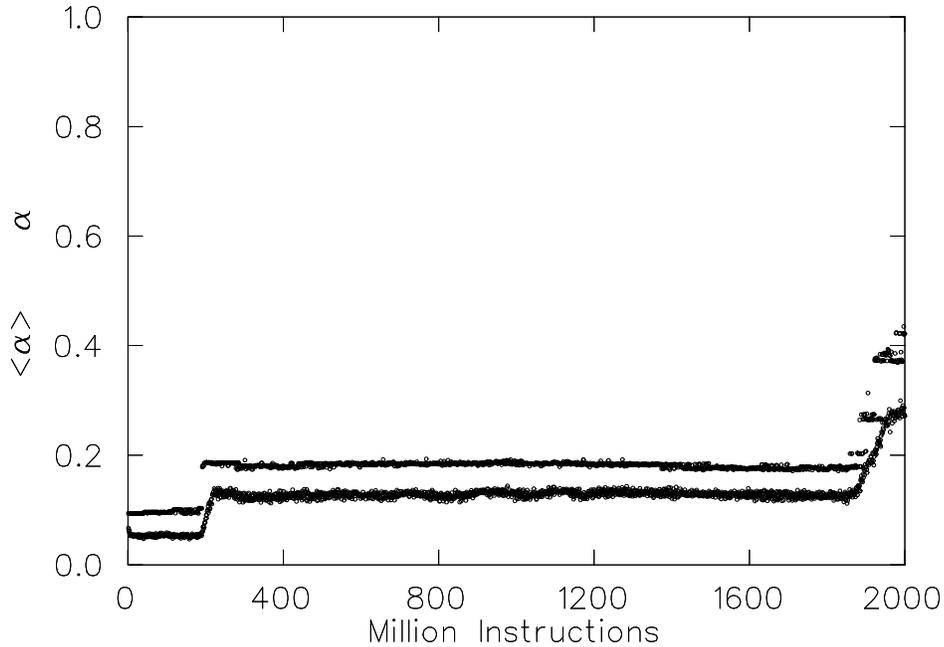

Fig. 3. Learning curve for a simulation with $R^{-1} = 150$ million instructions, same soup size as in Fig. 2. This run shows a long plateau at $\alpha \approx 0.2$ indicating trapping of the population in a meta-stable state.

The time it takes to escape such a state should be considerably reduced by choosing a larger soup size, which would allow for a more heterogenous population exploring different regions of the landscape at the same time. In tierra, the soup size (reserved memory space for cells) cannot easily be enlarged past a certain size, which entails that the population can equilibrate into a homogeneous phase rather easily. Consequently, it is important to investigate learning characteristics for different soup sizes. Preliminary studies have shown that increasing the soup size does enhance the learning fraction (fraction of runs that have learned before a certain cut-off time), and decreases the spread in learning times.

Tab. 3 shows the result for all 80 runs, used in Fig. 4, at soup size 131072 instructions[14]. The number of cells in such a soup is variable, but is of the order of magnitude of 800 cells of length 80.

| $R[10^{-8}]$ | 0.5 | 0.667 | 1.0 | 1.333 | 1.667 | 2.0 | 2.222 | 2.5 |
|---|---|---|---|---|---|---|---|---|
| 1 | 1586 | 843 | 242 | 453 | 1052 | 836 | e.c. | e.c. |
| 2 | 2802 | 1765 | 196 | 1601 | 606 | 1414 | e.c. | 397 |
| 3 | 1220 | 696 | 995 | 263 | >2002 | 240 | 293 | e.c. |
| 4 | 1413 | >3144 | 1201 | >2026 | 586 | 596 | >2022 | e.c. |
| 5 | 1136 | >2023 | 1041 | 407 | 406 | 507 | 343 | e.c. |
| 6 | >2025 | 1927 | >2019 | 357 | 380 | 270 | 809 | e.c. |
| 7 | >2090 | 1922 | 330 | 520 | 625 | 327 | 225 | e.c. |
| 8 | 901 | 1273 | 442 | 271 | 624 | 587 | e.c. | e.c. |
| 9 | 1353 | 1374 | >2251 | >2094 | >2117 | e.c. | e.c. | e.c. |
| 10 | 2252 | 1233 | 581 | >2087 | 406 | 222 | 803 | e.c. |

**Tab. 3** Learning times (in million instructions executed) for mutation rates from $0.5 \times 10^{-8}$ to $2.5 \times 10^{-8}$ for soup size $s = 131072$ instructions. An entry preceded by a "greater than" sign signifies that the task was not learned before that time and the run was interrupted. An entry "e.c." means that the population ceased to reproduce due to the "error catastrophe" as mentioned above.

Each column in Tab. 3 contains the learning times (defined as the time a genotype that successfully adds first dominates the population, i.e., has the most living copies in the soup), or in case adding was not achieved, the time the simulation was interrupted, preceded by a "greater than" symbol. Each run was taken to a minimum of 2 billion instructions executed. An entry "e.c." implies that the run died soon after inoculation due to the error-catastrophe.

---

[14]For technical reasons, the soup size has to be a power of 2.

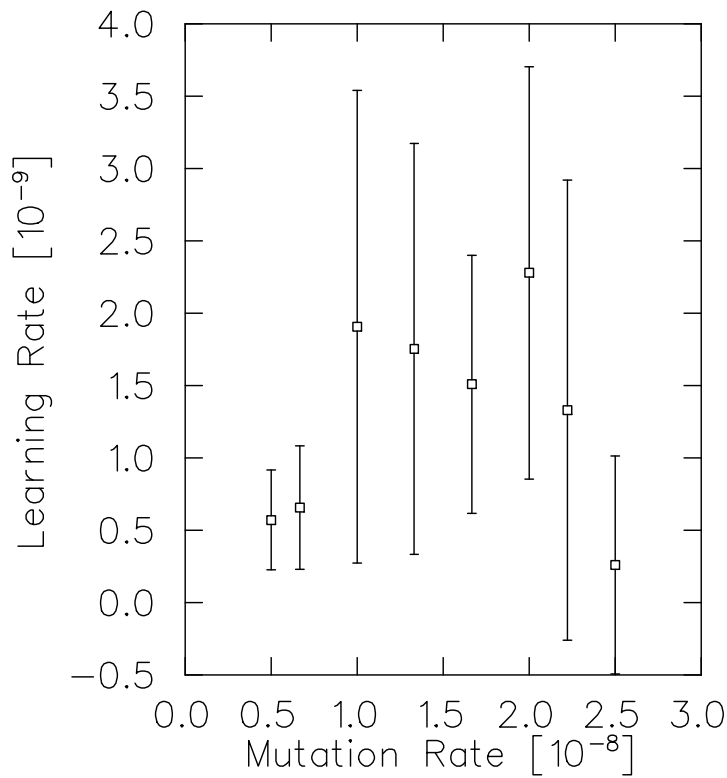

Fig. 4. Average learning rate as a statistical average of inverse learning times vs. mutation rate. The error bars are 1 $\sigma$ standard deviations.

Averaging the inverse learning times (learning rates) from each column of Tab. 5 yields Fig. 4[15]. As expected, the scattering of the data, implying large standard deviations, does not allow for definite conclusions on the behaviour of the learning rate. However, we can define a "learning fraction" by determining the fraction $f_X$ of runs at a given mutation rate that achieved learning at a time $t < t_c = X$, where $t_c$ is a cut-off that reflects the time-scale of learning in this environment for this task. As an example, the learning fraction with cut-off 1000 (million) at mutation rate $R = 1.0 \times 10^{-8}$ is $f_{1000}(1.0) = 0.6$ i.e., six out of ten runs resulted in a population that successfully added before 1000 million instructions were executed. This procedure allows us to obtain the curves presented in Fig. 5.

---

[15] Runs that did not achieve learning where given an infinite learning time, i.e., a learning rate of zero.

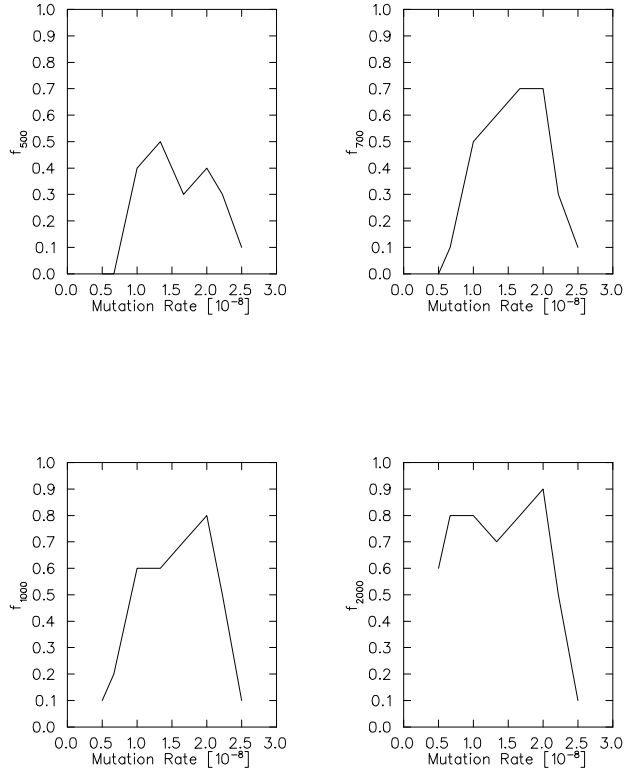

Fig. 5. Learning fractions $f_X$ as defined in the text vs. mutation rate for various cutoffs. (a) cutoff $X = 500$ million instructions, (b) $X = 700$, (c) $X = 1000$, and (d) $X = 2000$.

Clearly, choosing the cut-off scale too low would not reflect the learning characteristics of the soup, and neither would a high choice. In fact, this procedure implicitly determines simultaneously the window in mutation rate when learning is most effective, *and* an estimate for the time-scale of learning for this particular system and task.

Choosing the cut-off time scale (in units of million instructions executed) to be $t_c = 800 \pm 200$ the behaviour of the learning fraction suggests that learning becomes more and more effective as the mutation rate is increased up to a point where the soup dissolves as a result of the error catastrophe. This strongly suggests that evolutionary learning is most effective at the "edge of chaos" (see [13] for other examples of complex behaviour at the edge of chaos). The time-scale for learning determined here, however, is certainly not universal but depends on soup-size, initial creature, and bonus structure.

# 6 Conclusions

Evolutionary learning as displayed by the artificial system presented here has a number of fascinating characteristics that it may share with the natural genetic system that gave rise to bacterial DNA. As a learning system for practical applications tierra falls short in many respects, as must any auto-adaptive system at this stage. We have tried in this paper to extract universal characteristics of the auto-adaptive learning process, characteristics that should be reproducible by any software that incorporates the basic ingredients for auto-adaptive systems described earlier. Naturally, besides the universal characteristics there are features that must depend on the specifics of the implementation; it is the investigators task to isolate them. Particularly, the size and dimensionality of the tierran soup, i.e., the physical memory that the cells inhabit, has an influence on the global, and critical, behaviour of the population. In its present configuration, the tierran soup is one-dimensional, i.e., each cell has exactly two neighbours. This may be the most important limitation of tierra as it restricts growth and information transfer through the population, which determines equilibration times and self-organized behaviour. These aspects will be addressed for a two-dimensional genetic system in the near future [7].

**Acknowledgments**

I would like to thank Steve Koonin for the initial suggestion that lead to this work, as well as for continuing support and encouragement. This work was supported in part by NSF grant # PHY91-15574 and by a Caltech Divisional fellowship.